\documentclass[11pt,twoside,letterpaper]{article} 
\usepackage{times,fancyhdr}
\usepackage[dvips]{graphicx}
\usepackage{amsmath,epsfig,graphicx,amssymb}
\usepackage{listings}


\makeatletter 
\def\cleardoublepage{\clearpage\if@twoside \ifodd\c@page\else%
    \hbox{}%
    \thispagestyle{empty}%
    \newpage%
    \if@twocolumn\hbox{}\newpage\fi\fi\fi} 
\makeatother

\begin{document}
\title{
{\begin{flushleft}
\vskip 0.45in
{\normalsize\bfseries\textit{Chapter}}
\end{flushleft}
\vskip 0.45in
\bfseries\scshape Motion of Spinning Particles  around Black Holes}}
\author{\itshape Jose Miguel Ladino\thanks{E-mail address: jmladinom@unal.edu.co}\\
Universidad Nacional de Colombia. Sede Bogotá.\\ Facultad de Ciencias.Observatorio Astronómico Nacional.\\ 
Ciudad Universitaria. Bogotá, Colombia.\\
\\
\itshape Carlos Andrés del Valle\thanks{E-mail address: cdelv@unal.edu.co}\\
Universidad Nacional de Colombia. Sede Bogotá.\\ Facultad de Ciencias. Departamento de Física.\\ Ciudad Universitaria. Bogotá, Colombia.\\
\\
\itshape Eduard Larrañaga\thanks{E-mail address: ealarranaga@unal.edu.co}\\
Universidad Nacional de Colombia. Sede Bogotá.\\ Facultad de Ciencias. Observatorio Astronómico Nacional.\\ 
Ciudad Universitaria. Bogotá, Colombia.\\
 }
\date{}
\maketitle
\thispagestyle{empty}
\setcounter{page}{1}
\thispagestyle{fancy}
\fancyhead{}
\fancyhead[L]{In: Book Title \\ 
Editor: Editor Name, pp. {\thepage-\pageref{lastpage-01}}} 
\fancyhead[R]{ISBN 0000000000  \\
\copyright~2006 Nova Science Publishers, Inc.}
\fancyfoot{}
\renewcommand{\headrulewidth}{0pt}

\begin{abstract} 
The motion of spinning particles around compact objects, for example a rotating stellar object moving around a supermassive black hole, is described by differential equations that are, in general, non-integrable. In this work, we present a computational code that integrates the equations of motion of a spinning particle moving in the equatorial plane of a spherically symmetric spacetime and gives a visual representation of its trajectory. This code is open source, freely available and modular, so that users may extend its application not only to the Schwarzschild metric, but also to other backgrounds.
\end{abstract}

\vspace{2in}

\noindent  \noindent \textbf{Keywords:} Black Holes. Equations of Motion. Spinning Particles. 


\pagestyle{fancy}
\fancyhead{}
\fancyhead[EC]{J. M. Ladino, C. A. del Valle and E. Larrañaga}
\fancyhead[EL,OR]{\thepage}
\fancyhead[OC]{Motion of Spinning Particles  around Black Holes}
\fancyfoot{}
\renewcommand\headrulewidth{0.5pt} 
\section{Introduction}
\pagestyle{fancy}

Macroscopic astrophysical objects classically have a spin angular momentum associated with their rotation. When these objects, such as  stars or compact objects (black holes or neutron stars), belong to physical systems (e.g. binaries), it is well known that the existence of spin has consequences on the behavior of the dynamics of the system. A particular case of interest is that of spinning bodies orbiting around supermassive black holes and its relation with the production of gravitational waves. 

In order to introduce the classical spin to the description of the dynamics of extended bodies in curved spacetimes, Mathisson demonstrated in 1937 that the equations of motion show an interaction between the Riemann curvature tensor and the spin of the test particle \cite{mathisson1937zitternde}. Later, in 1951, Papapetrou made important contributions in the covariant treatment of the equations of motion, showing once again, that a spinning object moves in orbits that will differ from the geodesics \cite{1951RSPSA.209..248P}. Another significant input to this framework was made by Dixon in 1970, when he reformulated the equations of motion in a generalized form to hold exactly for an extended body \cite{1970RSPSA.314..499D}. Most of these approaches are restricted to the pole-dipole approximation, reducing the problem by considering that the body is not itself contributing to the gravitational field, and where just monopole and dipole or the mass and spin terms, respectively, are taken into account \cite{Toshmatov2019}. Finally, it has been studied the use of the so-called spin supplementary conditions, which fixes the center of the mass of the spinning body \cite{Lukes-Gerakopoulos:2014dma, Timogiannis2021,  Lukes-Gerakopoulos2017, Mikoczi2017, Costa2015}.
Following this logical path of results, the formalism that will be used in this work to describe the dynamics of a spinning particle in a curved spacetime is described by the Mathisson-Papapetrou-Dixon (MPD) equations.\\ 
\\
Although the motion of spinning particles in black holes backgrounds is, in general, a non-integrable problem, it has been analyzed using the MPD equations under different approaches. One of the first methods considers only the equatorial motion of a spinning particle in Schwarzschild and Kerr spacetimes, and obtains the behavior of some of the properties of the innermost stable circular orbit \cite{Suzuki1998}. Generalizations of this analysis, including the characteristics of black hole solutions from other gravitational theories, are presented in papers such as \cite{Tanaka1996, PhysRevD.97.084056, Zhang2019,Harms2017, Zhang2016,Pugliese2013, Conde2019, Jefremov2015, Nucamendi2020, Liu2020, An2018, Guo2016, Zhang(2)2019, Cariglia2012, Han2010}. 
The motion of spinning particles in static and spherically symmetric backgrounds have been also studied in literature, including the Schwarzschild–de Sitter spacetime \cite{Mortazavimanesh2009, Kunst2015,  Plyatsko2018}, the Reissner-Nordström-de Sitter black hole \cite{Stuchlik2001} and in general Schwarzschild-like space–times \cite{Mohseni2010}. Motion in backgrounds from alternative gravity theories was studied too. For example, in the background of a Schwarzschild black hole surrounded by quintessential matter field \cite{Sheoran2020},  in four-dimensional Einstein–Gauss–Bonnet space–time \cite{Zhang2020},  around a charged Hayward black hole \cite{Larranaga2020} or in a cosmological and general static spherically symmetric background \cite{Zalaquett2014}. All of these approaches prove that the spin of the test particle produce a non geodesic orbit, due to the existence of an additional force. 

There are currently various astrophysical systems of interest that have emerged in recent years, including the collision processes between spinning particles \cite{Maeda2018,Mukherjee2018, Zhang2018(2), Liu2020, Yuan2020, Hackmann2020, Liu2018, Armaza2016}, the study of black hole backgrounds as accelerators for spinning particles \cite{Sheoran2020, An2018, Zhang2016, Zaslavskii2016, Zhang(2)2019} or the emission of gravitational radiation from spinning particles orbiting around or plunging into a black hole \cite{Saijo1998, Mino1996, Harms2016, Han2010, Harms2017}. \\
\\
In this work, we provide a numerical code, which is publicly accessible and modifiable, to solve the equations of motion of a spinning particle orbiting around a spherically symmetric and static spacetime. Furthermore, it will give us the possibility of visualizing the trajectories based on the initial values of the system. This chapter is organized as follows: in Sec. 2, we present the MPD equations and the effects on the spin tensor when the motion is restricted to the equatorial plane.  In Sec. 3, we introduce the metric that describe a spherically symmetric and static background to determine the components of the spin tensor.  In Sec. 4, we obtain the components of the momentum vector from the associated conserved quantities of the motion. Then, the expressions of the radial and angular velocities are found in Sec. 5. Later, we show a brief summary of the step by step to solve the equations of motion computationally in Sec. 6. and finally, in Sec. 7, the code is commented.

\section{Equations of Motion for a Spinning Particle}
\pagestyle{fancy}

\subsection{The Mathisson-Papapetrou-Dixon Equations}

The MPD equations describe the motion of a spinning particle in a curved background. They state that \cite{mathisson1937zitternde,1951RSPSA.209..248P,1970RSPSA.314..499D}
\begin{align}
    \begin{cases}
    \frac{DP^\mu}{D\lambda} = &-\frac{1}{2} R^\mu_{\, \, \nu \alpha \beta} u^\nu S^{\alpha \beta}\\
    \frac{DS^{\mu \nu}}{D\lambda} = & P^\mu u^\nu - u^\mu P^\nu ,
    \end{cases} \label{eq:MPDequations}
\end{align}
where $P^\mu$ and $u^\mu = \frac{dx^\mu}{d\lambda}$ are the momentum and velocity vectors of the test particle, $R^\mu_{\, \, \nu \alpha \beta}$ is the Riemann tensor and $S^{\mu \nu}$ represents the spin tensor that defines the spin angular momentum of the test particle through the relation
\begin{equation}
    s^2 = \frac{1}{2} S^{\mu\nu} S_{\mu \nu}. 
\end{equation}

As is well known, the proper mass of the test particle is given by the norm of the momentum vector,
\begin{equation}
    m^2 = -P_\alpha P^\alpha. \label{eq:MomentumNormalization}
\end{equation}
but, introducing the quantity
\begin{equation}
    \mu = - P^\beta u_\beta,
\end{equation}
equations (\ref{eq:MPDequations}) gives the relation
\begin{equation}
    P^\alpha = \mu u^\alpha - u^\beta \frac{DS^{\alpha \beta}}{D\lambda},
\end{equation}
indicating that the momentum and the velocity vectors are no longer parallel for spinning particles. This fact gives the freedom to introduce an additional condition which may produce different trajectories \cite{PhysRevD.97.084056,Lukes-Gerakopoulos:2014dma}. There are many equations that we can introduce, but we will specifically use the \textit{Tulczyjew spin-supplementary condition} \cite{Hojman_2012,Lukes-Gerakopoulos:2014dma}, which restricts the spin tensor to generate only rotations through the relation
\begin{equation}
    P_\mu S^{\mu \nu} = 0. \label{eq:TulczjewCondition}
\end{equation}

Due to the spherical symmetry of the background, it is possible to consider, without loosing generality, that the orbital motion is restricted to the equatorial plane, $\theta = \frac{\pi}{2}$, which implies $P_2 = 0$. For simplicity, we will only consider the spin aligned or anti-aligned orbits. Thus, we impose the condition $S^{2\mu} = 0$ in equation (\ref{eq:TulczjewCondition}), to obtain the non-vanishing components of the spin tensor as
\begin{align}
    \begin{cases}
    S^{01} = -S^{10} = &\frac{P_3}{P_0} S^{13}\\
    S^{03} = -S^{30} = &- \frac{P_1}{P_0} S^{13}.
    \end{cases} \label{eq:SpinTensorComponents}
\end{align}

\section{Spherically Symmetric Static Spacetimes}

In this work, we will consider a spherically symmetric and static spacetime, with the following general line element
\begin{equation}
    ds^2 = g_{00} dt^2 + g_{11} dr^2 + g_{22}d\theta^2 + g_{33}d\phi^2.
\end{equation}

Using this metric, together with equations (\ref{eq:TulczjewCondition}) and (\ref{eq:SpinTensorComponents}), the spin angular momentum of the test particle is written as
\begin{align}
    s^2 = & \left[g_{00} g_{11}P_3^2 + g_{00} g_{33}P_1^2 + g_{11} g_{33}P_0^2 \right]\frac{(S^{13})^2}{P_0^2}.
\end{align}

The determinant of the metric tensor is given by
\begin{equation}
    g = \det g_{\mu \nu} = g_{00} g_{11} g_{22} g_{33}
\end{equation}
while the inverse tensor is calculated as
\begin{equation}
    g^{\mu \nu} = \frac{1}{g} 
    \begin{pmatrix}
    g_{11} g_{22} g_{33} & 0 & 0 & 0\\
    0 & g_{00} g_{22} g_{33} & 0 & 0\\
     0 & 0 & g_{00} g_{11} g_{33} & 0 \\
      0 & 0 & 0 & g_{00} g_{11} g_{22}
    \end{pmatrix}.
\end{equation}

Then, using these results we obtain
\begin{align}
    s^2 = & -m^2 (g_{00} g_{11} g_{33})\frac{(S^{13})^2}{P_0^2}.
\end{align}

Solving for the spin tensor component $S^{13}$ and using this result in equations (\ref{eq:SpinTensorComponents}) gives 
\begin{align}
\begin{cases}
    S^{13} = &-\frac{sP_0}{m} \frac{1}{\sqrt{-g_{00} g_{11} g_{33}}}\\
    S^{01} = &-\frac{sP_3}{m} \frac{1}{\sqrt{-g_{00} g_{11} g_{33}}}\\
    S^{03} = &-\frac{sP_1}{m} \frac{1}{\sqrt{-g_{00} g_{11} g_{33}}}.
\end{cases}\label{eq:SpinTensorComponents2}
\end{align}

\section{Conserved Quantities in the Motion of a Spinning Test Particle}

Given the existence of a Killing vector $k$, we associate the existence of a conserved quantity in the motion of the spinning particle through the relation
\begin{equation}
    C_k = P^\mu k_\mu + \frac{1}{2} S^{\mu \nu} \nabla_\mu k_\nu .
\end{equation}

Assuming that the metric tensor components do not depend explicitly on the coordinates $t$ and $\phi$, there are two Killing vectors,
\begin{equation}
    \xi = \partial_t \hspace{1cm} \text{ and } \hspace{1cm} \zeta = \partial_\phi ,   
\end{equation}
corresponding to the conservation of energy and angular momentum, respectively. For the first of these vectors we have
\begin{align}
    C_t = -E = & \xi^\mu P_\mu + \frac{1}{2} S^{\mu \nu} \nabla_\mu \xi_\nu.
\end{align}

The covariant derivative in the second term gives
\begin{align}
    S^{\mu \nu} \nabla_\mu \xi_\nu = S^{\mu \nu} \partial_\mu \xi_\nu - S^{\mu \nu} \Gamma^\alpha _{\mu \nu} \xi_\alpha = S^{\mu \nu} \partial_\mu \xi_\nu
\end{align}
and because $\xi_\mu = g_{\mu\nu} \xi^\nu = g_{00} \delta^0_\mu$, we obtain  
\begin{align}
    E = & -P_0 - \frac{1}{2} S^{\mu 0} \partial_\mu g_{00}.
\end{align}

Considering the dependence $g_{00} = g_{00}(r)$, we get
\begin{equation}
    E = -P_0 - \frac{1}{2} S^{1 0} \partial_r g_{00}
\end{equation}
and using equations (\ref{eq:SpinTensorComponents2}) this becomes
\begin{equation}
    E = -P_0 - \frac{1}{2} \frac{sP_3}{m} \frac{\partial_r g_{00}}{\gamma} , \label{eq:Energy}
\end{equation}
where we introduce $\gamma = \sqrt{-g_{00} g_{11} g_{33}}$. A similar procedure for the Killing vector $\zeta$ gives
\begin{align}
    C_\phi = J = & \zeta^\mu P_\mu + \frac{1}{2} S^{\mu \nu} \nabla_\mu \zeta_\nu = \zeta^\mu P_\mu + \frac{1}{2} S^{\mu \nu} \partial_\mu \zeta_\nu.
\end{align}

This time, the Killing vector satisfies $\zeta_\mu = g_{\mu\nu} \zeta^\nu = g_{33} \delta^3_\mu$, and therefore 
\begin{align}
    J = & P_3 + \frac{1}{2} S^{\mu 3} \partial_\mu g_{33}.
\end{align}

Considering now the dependence $g_{33} = g_{33}(r)$ ( because considering that the motion occurs in the equatorial plane implies that there is no dependence with the coordinate $\theta$ ), we obtain
\begin{equation}
    J = P_3 + \frac{1}{2} S^{1 3} \partial_r g_{33}
\end{equation}
and using equations (\ref{eq:SpinTensorComponents2}) this becomes
\begin{equation}
    J = P_3 - \frac{1}{2} \frac{sP_0}{m} \frac{\partial_r g_{33}}{\gamma} . \label{eq:AngularMomentum}
\end{equation}

From equations (\ref{eq:Energy}) and (\ref{eq:AngularMomentum}) we solve to obtain the components $P_0$ and $P_3$ of the momentum vector as
\begin{equation}
    \begin{cases}
    P_0 = & - \frac{2m\gamma \left( 2m\gamma E + s J \partial_r g_{00} \right)}{4 m^2 \gamma^2 + s^2 \partial_r g_{00} \partial_r g_{33}} \\
    P_3 = &  \frac{2m\gamma \left( 2m\gamma J + s E \partial_r g_{33} \right)}{4 m^2 \gamma^2 + s^2 \partial_r g_{00} \partial_r g_{33}}.
    \end{cases}
\end{equation}

The component $P_1$ is obtained from the normalization condition (\ref{eq:MomentumNormalization}), giving
\begin{equation}
    (P_1)^2 = - \frac{m^2 + g^{00} (P_0)^2 + g^{33}(P_3)^2}{g^{11}}.
\end{equation}

\section{The Equations of Motion revisited}

Now, we are able to write explicitly the equations of motion given in (\ref{eq:MPDequations}). Using a parameter $\lambda$ for which $\dot{u}^0 = 1$, we have that the equation for the component $S^{01}$ of the spin tensor is
\begin{equation}
    \frac{DS^{01}}{D\lambda} = P^0 u^1 - P^1.
\end{equation}

Using equations (\ref{eq:SpinTensorComponents}) and (\ref{eq:SpinTensorComponents2}) in the left hand side gives
\begin{equation}
    -\frac{s P_3}{m}\frac{D}{D\lambda} \left( \frac{1}{\gamma} \right) -\frac{s}{m\gamma} \frac{DP_3}{D\lambda}  = P^0 \dot{r} - P^1.
\end{equation}
Since the factor $\frac{1}{\gamma} = \frac{1}{\sqrt{-g_{00}g_{11}g_{33}}}$ depends only on the coordinate $r$, we have that
\begin{equation}
    \frac{D}{D\lambda} \left( \frac{1}{\gamma} \right) = \frac{dx^\mu}{d\lambda} \nabla_\mu \left( \frac{1}{\gamma} \right) = \frac{dr}{d\lambda} \partial_r \left( \frac{1}{\gamma} \right) = -\frac{\dot{r}}{\gamma^2} \partial_r \gamma .
\end{equation}

Replacing in the equation of motion gives
\begin{equation}
     \dot{r} \left[  P^0 - \frac{s P_3}{m\gamma^2} \partial_r \gamma \right] = P^1 -\frac{s }{m\gamma} \frac{DP_3}{D\lambda}.
\end{equation}

The last term in the right hand side can be evaluated using the equation of motion for $P_3$ arising from (\ref{eq:MPDequations}),
\begin{equation}
    \frac{DP_3}{D\lambda} = -\frac{1}{2} R_{3 \nu \alpha \beta} u^\nu S^{\alpha \beta}.
\end{equation}

This gives the final result
\begin{equation}
     \dot{r} \left[  P^0 - \frac{s P_3}{m\gamma^2}  \partial_r \gamma \right] = P^1 +\frac{s }{2m\gamma} R_{3 \nu \alpha \beta} u^\nu S^{\alpha \beta}.
\end{equation}

A similar procedure using the equation of motion for the component $S^{03}$ of the spin tensor, gives the following relation,
\begin{equation}
     \dot{\phi} P^0 + \frac{s P_1}{m\gamma^2 } \dot{r} \partial_r \gamma  = P^3 -\frac{s}{2m\gamma}  R_{1 \nu \alpha \beta} u^\nu S^{\alpha \beta}.
\end{equation}

Note that using the anti-symmetry properties (in the first two indices) of the Riemann tensor and the fact that the particle is moving on the equatorial plane $\dot{\theta} = 0$, we have that
\begin{equation}
    R_{3 \nu \alpha \beta} u^\nu S^{\alpha \beta} = R_{3 0 \alpha \beta}  S^{\alpha \beta} + R_{3 1 \alpha \beta} \dot{r} S^{\alpha \beta}
\end{equation}
and
\begin{equation}
    R_{1 \nu \alpha \beta} u^\nu S^{\alpha \beta} = R_{1 0 \alpha \beta}  S^{\alpha \beta} + R_{1 3 \alpha \beta} \dot{\phi} S^{\alpha \beta}.
\end{equation}

Therefore, the equations of motion are now

\begin{equation}
    \begin{cases}
    \dot{r} \left[  P^0 - \frac{s P_3}{m\gamma^2} \partial_r \gamma \right] = &P^1 +\frac{s }{2m\gamma} R_{3 0 \alpha \beta}  S^{\alpha \beta}  + \frac{s }{2m\gamma} R_{3 1 \alpha \beta} S^{\alpha \beta} \dot{r} \\
    \dot{\phi} P^0 + \frac{s P_1}{m\gamma^2} \dot{r} \partial_r \gamma  = &P^3 -\frac{s}{2m\gamma} R_{1 0 \alpha \beta}  S^{\alpha \beta}  - \frac{s}{2m\gamma} R_{1 3 \alpha \beta}S^{\alpha \beta}   \dot{\phi} .
    \end{cases}
\end{equation}

From the first of these relations, we obtain an expression for the radial velocity, $\dot{r}$,
\begin{equation}
    \dot{r}  = \frac{P^1 +\frac{s }{2m\gamma} R_{3 0 \alpha \beta}  S^{\alpha \beta} }{  P^0 - \frac{s P_3}{m\gamma^2} \partial_r \gamma - \frac{s }{2m\gamma} R_{3 1 \alpha \beta} S^{\alpha \beta} } ,
\end{equation}
while the second relation gives an equation for the angular velocity $\dot{\phi}$,
\begin{equation}
    \dot{\phi} = \frac{P^3 - \frac{s P_1}{m\gamma^2} \dot{r} \partial_r \gamma -\frac{s}{2m\gamma} R_{1 0 \alpha \beta}  S^{\alpha \beta} }{P^0 + \frac{s}{2m\gamma} R_{1 3 \alpha \beta}S^{\alpha \beta}} .
\end{equation}

\section{Summary of the Equations and Procedure to Solve Them Computationally}

In order to solve the equations of motion for the particular case of the spinning particle in the equatorial plane of a spherically symmetric and static spacetime, we will impose initial values for the following quantities:

\begin{itemize}
    \item Proper mass of the particle: $m$ (conserved)
    \item Initial position: $t_0$, $r_0$ and $\phi_0$ 
    \item Total Energy of the particle: $E$ (conserved)
    \item Orbital Angular Momentum : $j$ (conserved)
    \item Spin Angular Momentum : $s$ (conserved)
\end{itemize}

The total angular momentum is determined by the values of $j$ and $s$ by
\begin{equation}
    J = j + s.
\end{equation}

Given these initial values, we proceed with the following steps:
\begin{enumerate}
    \item Obtain the initial values for the components of the momentum using the equations
\begin{equation}
    \begin{cases}
    P_0 = & - \frac{2m\gamma \left( 2m\gamma E + s J \partial_r g_{00} \right)}{4 m^2 \gamma^2 + s^2 \partial_r g_{00} \partial_r g_{33}} \\
    P_3 = &  \frac{2m\gamma \left( 2m\gamma J + s E \partial_r g_{33} \right)}{4 m^2 \gamma^2 + s^2 \partial_r g_{00} \partial_r g_{33}}\\
    (P_1)^2 = & - \frac{m^2 + g^{00} (P_0)^2 + g^{33}(P_3)^2}{g^{11}}.
    \end{cases}
\end{equation}

\item Determine the initial values of the spin tensor components
\begin{align}
\begin{cases}
    S^{13} = &-\frac{sP_0}{m\gamma} \\
    S^{01} = &-\frac{sP_3}{m\gamma} \\
    S^{03} = &-\frac{sP_1}{m\gamma} 
\end{cases}.
\end{align}

\item Calculate the Riemann tensor components $R_{\mu \nu \alpha \beta}$. 

\item Numerically solve the equations 
\begin{equation}
    \dot{r}  = \frac{P^1 +\frac{s }{2m\gamma} R_{3 0 \alpha \beta}  S^{\alpha \beta} }{  P^0 - \frac{s P_3}{m\gamma^2} \partial_r \gamma - \frac{s }{2m\gamma} R_{3 1 \alpha \beta} S^{\alpha \beta} } 
\end{equation}
and
\begin{equation}
    \dot{\phi} = \frac{P^3 - \frac{s P_1}{m\gamma^2} \dot{r} \partial_r \gamma -\frac{s}{2m\gamma} R_{1 0 \alpha \beta}  S^{\alpha \beta} }{P^0 + \frac{s}{2m\gamma} R_{1 3 \alpha \beta}S^{\alpha \beta}} 
\end{equation}
to obtain the new values of the coordinates $r$ and $\phi$. 
\item Begin the process again at step 1.
\end{enumerate}

\section{The MSPBH Code}

The MSPBH (Motion of Spinning Particles around Black Holes) Code\footnote{The MSPBH Code is available at https://github.com/cdelv/MSPBH} is a Python implementation that solves the equations of motion described above to obtain the trajectory of a spinning particle around a spherically symmetric compact object described by a metric tensor. The program uses symbolic algebra packages \lstinline{SymPy} and \lstinline{EinsteinPy} to calculate the required quantities, such as partial derivatives of the metric and the Riemann tensor, and the integration of the equations of motion is performed using \lstinline{odeint} from the \lstinline{SciPy} package. The code reads all the necessary variables, constants, initial conditions, and information about the metric needed to calculate the trajectories from a configuration file. When executing the code without configuration, a file called \lstinline{template.txt} is created with a usage example corresponding to the last stable orbit in the Schwarzschild metric. 

\begin{figure}[htb!]
    \centering
    \includegraphics[scale=0.45]{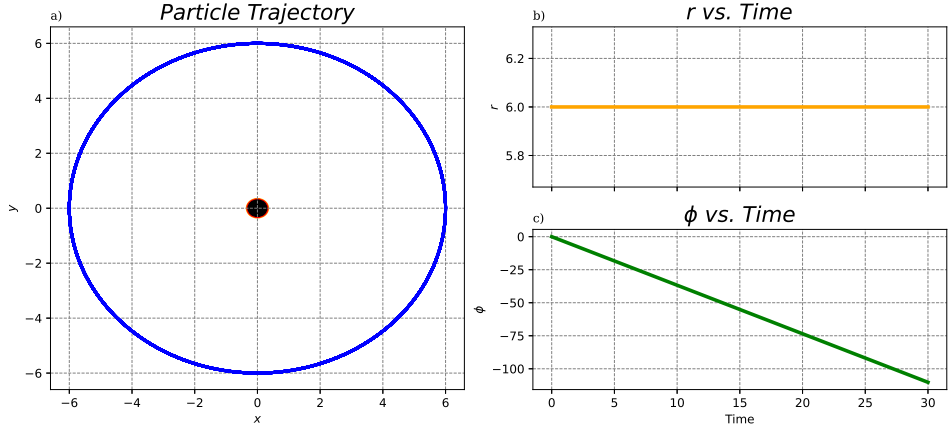}
    \vspace{30 mm}
    \caption{Trajectory of a particle with mass $m=1$, around a black hole with mass $M=1$. The initial conditions are $r_0=6$, $\phi_0=0$, the particle energy is $E=-\sqrt{8/9}$, the angular momentum is $j=\sqrt{12}$ and the spin is $s=0$.}
    \label{fig::trajectory1}
\end{figure}

Some of the parameters and options in the configuration file include the definition of the metric itself. For example, if the options \lstinline{User_tensor} and \lstinline{User_metric} are set to \lstinline{FALSE}, the program will use the incorporated Schwarzschild metric from the library \lstinline{Einsteinpy}, and compute the Riemann tensor. When \lstinline{User_tensor} is \lstinline{TRUE}, the program will use the components of the tensors given by the user on the configuration file (this option has precedence over \lstinline{User_metric}. In fact, when \lstinline{User_metric} is \lstinline{TRUE} and \lstinline{User_tensor} is \lstinline{FALSE}, the program will use the function \lstinline{Create_User_Metric_Tensor} to create a custom metric tensor with the library \lstinline{Einsteinpy}.
If \lstinline{User_metric} is False, you don't have to give the Riemann and metric tensor components on the template. In this case, once the metric tensor is determined, the code will compute the Riemann tensor using the \lstinline{EinsteinPy} library and proceeds to calculate the orbit. 
\begin{figure}[htb!]
    \centering
    \includegraphics[scale=0.45]{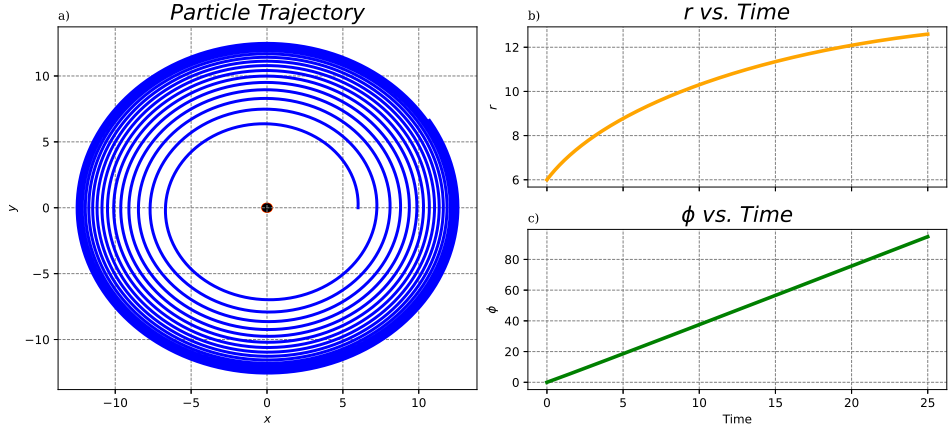}
    \vspace{30 mm}
    \caption{Trajectory of a particle with mass $m=1$, around a black hole with mass $M=1$. The initial conditions are $r_0=6$, $\phi_0=0$, the particle energy is $E=-\sqrt{8/9}$, the angular momentum is $j=\sqrt{12}$ and the spin is $s=3$.}
    \label{fig::trajectory2}
\end{figure}

A successful execution of the code will provide a series of plots (is the user sets options \lstinline{Plot_trajectoryy} and/or \lstinline{Plot_energy} to \lstinline{TRUE}), including the trajectory of the particle and the behavior of the coordinates $r(t)$ and $\phi (t)$, and two output files: \lstinline{'log.out'} and \lstinline{'data.csv'}. The first one has information about all the variables read from the configuration file, while the second is a CSV file where you will find the information about time, $r$ and $\phi$ coordinates, energy, and angular momentum of the particle at each step in the integration. In Figure \ref{fig::trajectory1}, the trajectory and the evolution of the coordinates of a non spinning particle can be seen.

For a spinning particle there are two possible configurations: the spin aligns with the angular momentum of the orbit or not. On Figure \ref{fig::trajectory2} we show the trajectory of the same particle described in Figure \ref{fig::trajectory1} but this time with a non-zero spin aligned with the angular momentum. 

\section{Conclusion}
We have considered the equations of motion of a spinning particle moving in the equatorial plane of a spherically symmetric and static background to write the expressions for the radial and angular velocities in terms of the components of the momentum vector,  the spin tensor and the Riemann tensor. After given a brief summary of the step by step algorithm to solve these equations of motion numerically, we have presented a Python implementation named the MSPBH Code, which solves numerically the equations of motion and provides a visualization of the trajectory. The code also gives the coordinates of the spinning particle together with the corresponding energy and angular momentum of the spinning particle.  

We have illustrated the implementation by considering two examples (non spinning particle and spinning particle moving around a Schwarzschild black hole) and showing the obtained trajectory. The results confirm that the spin of the particle causes a change in the trajectory, producing a fall into the black hole.

The MSPBH Code provides a practical and visual tool to demonstrate how the spin of the particle affects its own dynamics in a spherically symmetric and static spacetime. It is open source, freely available and modular, so that users may extend its application not only to the default schwarzschild metric, but also to other backgrounds included in the \lstinline{EinsteinPy} library or even by defining its own metric. \\

\emph{Acknowledgements}. This work was supported by the Universidad
Nacional de Colombia. Research Incubator No. 64-Computational Astrophysics and  Hermes Grant Code 41673.

\label{lastpage-01}


\begin{thebibliography}{0}    


 \bibitem{mathisson1937zitternde} M Mathisson. Das zitternde elektron und seine dynamik. Acta Phys. Pol, 6:218,
1937.
\bibitem{1951RSPSA.209..248P} A. Papapetrou. Spinning Test-Particles in General Relativity. I. Proceedings
of the Royal Society of London Series A, 209(1097):248–258, October 1951.

\bibitem{1970RSPSA.314..499D} W. G. Dixon. Dynamics of Extended Bodies in General Relativity. I. Momen-
tum and Angular Momentum. Proceedings of the Royal Society of London
Series A, 314(1519):499–527, January 1970.

\bibitem{Toshmatov2019} Bobir Toshmatov and Daniele Malafarina. Spinning test particles in the $\ensuremath{\gamma}$ 
spacetime. Phys. Rev. D, 100:104052, Nov 2019.

\bibitem{Lukes-Gerakopoulos:2014dma} Georgios Lukes-Gerakopoulos, Jonathan Seyrich, and Daniela Kunst. Investi-
gating spinning test particles: spin supplementary conditions and the Hamil-
tonian formalism. Phys. Rev. D, 90(10):104019, 2014.

\bibitem{Timogiannis2021} Iason Timogiannis, Georgios Lukes-Gerakopoulos, and Theocharis A. Aposto-
latos. Spinning test body orbiting around a schwarzschild black hole: Com-
paring spin supplementary conditions for circular equatorial orbits. Physical
Review D, 104(2), Jul 2021.

\bibitem{Lukes-Gerakopoulos2017} Georgios Lukes-Gerakopoulos. Time parameterizations and spin supplemen-
tary conditions of the mathisson-papapetrou-dixon equations. Phys. Rev. D,
96:104023, Nov 2017.

\bibitem{Mikoczi2017} Balázs Mikóczi. Spin supplementary conditions for spinning compact binaries.
Phys. Rev. D, 95:064023, Mar 2017.

\bibitem{Costa2015} L. Filipe O. Costa and José Natário. Center of Mass, Spin Supplementary
Conditions, and the Momentum of Spinning Particles, pages 215–258. Springer
International Publishing, Cham, 2015.

\bibitem{Suzuki1998} Shingo Suzuki and Kei-ichi Maeda. Innermost stable circular orbit of a spinning
particle in kerr spacetime. Physical Review D, 58(2), Jun 1998.

\bibitem{Tanaka1996} Takahiro Tanaka, Yasushi Mino, Misao Sasaki, and Masaru Shibata. Gravita-
tional waves from a spinning particle in circular orbits around a rotating black
hole. Phys. Rev. D, 54:3762–3777, Sep 1996.

\bibitem{PhysRevD.97.084056} Yu-Peng Zhang, Shao-Wen Wei, Wen-Di Guo, Tao-Tao Sui, and Yu-Xiao Liu.
Innermost stable circular orbit of spinning particle in charged spinning black
hole background. Phys. Rev. D, 97:084056, Apr 2018.

\bibitem{Zhang2019} Ming Zhang and Wen-Biao Liu. Innermost stable circular orbits of charged
spinning test particles. Physics Letters B, 789:393–398, 2019.

\bibitem{Harms2017} Georgios Lukes-Gerakopoulos, Enno Harms, Sebastiano Bernuzzi, and Alessan-
dro Nagar. Spinning test body orbiting around a kerr black hole: Circular
dynamics and gravitational-wave fluxes. Physical Review D, 96(6), Sep 2017.

\bibitem{Zhang2016} Yu-Peng Zhang, Bao-Min Gu, Shao-Wen Wei, Jie Yang, and Yu-Xiao Liu.
Charged spinning black holes as accelerators of spinning particles. Phys. Rev.
D, 94:124017, Dec 2016.

\bibitem{Pugliese2013} Daniela Pugliese, Hernando Quevedo, and Remo Ruﬀini. Equatorial circular
orbits of neutral test particles in the kerr-newman spacetime. Phys. Rev. D,
88:024042, Jul 2013.

\bibitem{Conde2019} Carlos Conde, Cristian Galvis, and Eduard Larrañaga. Properties of the inner-
most stable circular orbit of a spinning particle moving in a rotating maxwell-
dilaton black hole background. Phys. Rev. D, 99:104059, May 2019.

\bibitem{Jefremov2015} Paul I. Jefremov, Oleg Yu. Tsupko, and Gennady S. Bisnovatyi-Kogan. Inner-
most stable circular orbits of spinning test particles in schwarzschild and kerr
space-times. Phys. Rev. D, 91:124030, Jun 2015.

\bibitem{Nucamendi2020} Ulises Nucamendi, Ricardo Becerril, and Pankaj Sheoran. Bounds on spinning
particles in their innermost stable circular orbits around rotating braneworld
black hole. The European Physical Journal C, 80(1):35, Jan 2020.

\bibitem{Liu2020} Yunlong Liu and Xiangdong Zhang. Maximal eﬀiciency of the collisional penrose
process with spinning particles in kerr-sen black hole. The European Physical
Journal C, 80(1):31, Jan 2020.

\bibitem{An2018} Jincheng An, Jun Peng, Yan Liu, and Xing-Hui Feng. Kerr-sen black hole as
accelerator for spinning particles. Phys. Rev. D, 97:024003, Jan 2018.

\bibitem{Guo2016} Minyong Guo and Sijie Gao. Kerr black holes as accelerators of spinning test
particles. Physical Review D, 93(8), Apr 2016.

\bibitem{Zhang(2)2019} Songming Zhang, Yunlong Liu, and Xiangdong Zhang. Kerr–de sitter and kerr–
anti–de sitter black holes as accelerators for spinning particles. Phys. Rev. D,
99:064022, Mar 2019.

\bibitem{Cariglia2012} David Kubizňák and Marco Cariglia. Integrability of spinning particle mo-
tion in higher-dimensional rotating black hole spacetimes. Phys. Rev. Lett.,
108:051104, Jan 2012.

\bibitem{Han2010} Wen-Biao Han. Gravitational radiation from a spinning compact object around
a supermassive kerr black hole in circular orbit. Phys. Rev. D, 82:084013, Oct
2010.

\bibitem{Mortazavimanesh2009} M. Mortazavimanesh and Morteza Mohseni. Spinning particles in
schwarzschild–de sitter space–time. General Relativity and Gravitation,
41(11):2697–2706, Nov 2009.

\bibitem{Kunst2015} Daniela Kunst, Volker Perlick, and Claus Lämmerzahl. Isofrequency pair-
ing of spinning particles in schwarzschild–de sitter spacetime. Phys. Rev. D,
92:024029, Jul 2015.

\bibitem{Plyatsko2018} Roman Plyatsko, Volodymyr Panat, and Mykola Fenyk. Nonequatorial circular
orbits of spinning particles in the schwarzschild–de sitter background. General
Relativity and Gravitation, 50(11):150, Oct 2018.

\bibitem{Stuchlik2001} Zden ěk Stuchlík and Stanislav Hledík. Equilibrium of a charged spinning
test particle in reissner-nordström backgrounds with a nonzero cosmological
constant. Phys. Rev. D, 64:104016, Oct 2001.

\bibitem{Mohseni2010} Morteza Mohseni. Stability of circular orbits of spinning particles
in schwarzschild-like space–times. General Relativity and Gravitation,
42(10):2477–2490, Oct 2010.

\bibitem{Sheoran2020} Pankaj Sheoran, Hemwati Nandan, Eva Hackmann, Ulises Nucamendi, and
Amare Abebe. Schwarzschild black hole surrounded by quintessential matter
field as an accelerator for spinning particles. Phys. Rev. D, 102:064046, Sep
2020.

\bibitem{Zhang2020} Yu-Peng Zhang, Shao-Wen Wei, and Yu-Xiao Liu. Spinning test particle in
four-dimensional einstein–gauss–bonnet black holes. Universe, 6(8), 2020.

\bibitem{Larranaga2020} Eduard Larrañaga. Circular motion and the innermost stable circular orbit for
spinning particles around a charged hayward black hole background. Interna-
tional Journal of Modern Physics D, 29(16):2050121, 2020.

\bibitem{Zalaquett2014} Nicolas Zalaquett, Sergio A Hojman, and Felipe A Asenjo. Spinning massive
test particles in cosmological and general static spherically symmetric space-
times. Classical and Quantum Gravity, 31(8):085011, apr 2014.

\bibitem{Maeda2018} Kei-ichi Maeda, Kazumasa Okabayashi, and Hirotada Okawa. Maximal eﬀi-
ciency of the collisional penrose process with spinning particles. Phys. Rev. D,
98:064027, Sep 2018.

\bibitem{Mukherjee2018} Sajal Mukherjee. Collisional penrose process with spinning particles. Physics
Letters B, 778:54–59, 2018.

\bibitem{Zhang2018(2)} Ming Zhang, Jie Jiang, Yan Liu, and Wen-Biao Liu. Collisional penrose process
of charged spinning particles. Phys. Rev. D, 98:044006, Aug 2018.

\bibitem{Yuan2020} Xulong Yuan, Yunlong Liu, and Xiangdong Zhang. Collision of spinning parti-
cles near BTZ black holes. Chinese Physics C, 44(6):065104, jun 2020.

\bibitem{Hackmann2020} Eva Hackmann, Hemwati Nandan, and Pankaj Sheoran. Particle collisions near
static spherically symmetric black holes. Physics Letters B, 810:135850, 2020.

\bibitem{Liu2018} Yan Liu and Wen-Biao Liu. Energy extraction of a spinning particle via the su-
per penrose process from an extremal kerr black hole. Phys. Rev. D, 97:064024,
Mar 2018.

\bibitem{Armaza2016} Cristóbal Armaza, Máximo Banados, and Benjamin Koch. Collisions of spin-
ning massive particles in a schwarzschild background. Classical and Quantum
Gravity, 33(10):105014, apr 2016.

\bibitem{Zaslavskii2016} O. B. Zaslavskii. Schwarzschild black hole as particle accelerator of spinning
particles. EPL (Europhysics Letters), 114(3):30003, may 2016.

\bibitem{Saijo1998} Motoyuki Saijo, Kei-ichi Maeda, Masaru Shibata, and Yasushi Mino. Gravi-
tational waves from a spinning particle plunging into a kerr black hole. Phys.
Rev. D, 58:064005, Aug 1998.

\bibitem{Mino1996} Yasushi Mino, Masaru Shibata, and Takahiro Tanaka. Gravitational waves
induced by a spinning particle falling into a rotating black hole. Phys. Rev. D,
53:622–634, Jan 1996.

\bibitem{Harms2016} Enno Harms, Georgios Lukes-Gerakopoulos, Sebastiano Bernuzzi, and Alessan-
dro Nagar. Asymptotic gravitational wave fluxes from a spinning particle in
circular equatorial orbits around a rotating black hole. Phys. Rev. D, 93:044015,
Feb 2016.

\bibitem{Hojman_2012} Sergio A Hojman and Felipe A Asenjo. Can gravitation accelerate neutrinos?
Classical and Quantum Gravity, 30(2):025008, dec 2012

\end{thebibliography}
\end{document}